\documentclass[vecphys]{svmult}
\usepackage[bottom]{footmisc}
\usepackage{cite,slashed,amssymb}
\usepackage{graphicx}
\usepackage{ifpdf}
\ifx\pdfoutput\undefined
   \pdffalse
   \usepackage{cite}
 \else
   \pdfoutput=1
   \pdftrue
  \usepackage[pdftex]{hyperref}
\fi
  \DeclareGraphicsExtensions{.pdf}
\newcommand{\Red}[1]{#1}
\newcommand{\Blue}[1]{#1}
\newcommand{\ft}[2]{{\textstyle\frac{#1}{#2}}}
\def\Tr{\mathop{\rm Tr}\nolimits}
\def\rmi{{\rm i}}
\def\rmd{{\rm d}}

\def\Re{\mathop{\rm Re}\nolimits}
\def\Im{\mathop{\rm Im}\nolimits}
\def\cn{{\cal N}} 

\newsavebox{\uuunit}
\sbox{\uuunit}
    {\setlength{\unitlength}{0.825em}
     \begin{picture}(0.6,0.7)
        \thinlines
        \put(0,0){\line(1,0){0.5}}
        \put(0.15,0){\line(0,1){0.7}}
        \put(0.35,0){\line(0,1){0.8}}
       \multiput(0.3,0.8)(-0.04,-0.02){12}{\rule{0.5pt}{0.5pt}}
     \end {picture}}
\newcommand {\unity}{\mathord{\!\usebox{\uuunit}}}

\newcommand{\SU}{\mathop{\rm SU}}
\newcommand{\SO}{\mathop{\rm SO}}
\newcommand{\U}{\mathop{\rm {}U}}
\newcommand{\Sp}{\mathop{\rm Sp}}
\newcommand{\soa}{\mathop{\mathfrak{so}}}   

\newcommand{\USp}{\mathop{\rm USp}}
\newcommand{\OSp}{\mathop{\rm OSp}}

\newcommand{\begintable}[3]
  {\begin{table}\centering \caption{#1}
  \label{#2} #3\end{table}}
\newcommand{\beginlstable}[3]
  {\landscape\begin{table}\entering \caption{#1} \label{#2} #3\end{table}\endlandscape}

\begin{document}\pagenumbering{gobble}
\begin{flushright}
arXiv:1306.XXX
\end{flushright}
\vspace{.5cm}
\begin{center}
\baselineskip=16pt {\LARGE   Superconformal Symmetry and Higher-Derivative
Lagrangians
}\\
\vfill
{\large Antoine Van Proeyen
  } \\
\vfill
{\small Instituut voor Theoretische Fysica, Katholieke Universiteit Leuven,\\
       Celestijnenlaan 200D B-3001 Leuven, Belgium.
\\ \vspace{6pt}
 }
\end{center}
\vfill
\begin{center}
{\bf Abstract}
\end{center}
{\small Superconformal methods are useful to build invariant actions in
supergravity. We have a good insight in the possibilities of actions that
are at most quadratic in spacetime derivatives, but insight in general
higher-derivative actions is missing. Recently higher-derivative actions
got more attention for several applications. One of these is the
understanding of finiteness of loop computations in supergravities.
Divergences can only occur if invariant counterterms or anomalies exist.
One can wonder whether conformal symmetry might also play a role in this
context. In order to construct higher-derivative supergravities with the
conformal methods, one should first get more insight in such rigid
supersymmetric actions with extra fermionic symmetries. We show how
Dirac--Born--Infeld actions with Volkov--Akulov supersymmetries can be
constructed in all orders.}
 \vfill

\hrule width 3.cm \vspace{2mm}{\footnotesize \noindent Contribution to the
proceedings of 'Breaking of supersymmetry and Ultraviolet  Divergences in
extended Supergravity', Frascati, March 2013, to be published as Springer
Lecture Notes
\\
e-mail: \texttt{Antoine.VanProeyen@fys.kuleuven.be}}

\newpage

\pagenumbering{arabic}

\section{Introduction}
 \label{sec:intro}
In the last 35 years, supergravity actions with terms that are at most
quadratic in spacetime derivatives have been studied a lot. But recently
higher-derivative terms in supergravity actions got more interest. There
are different reasons for this. They appear as order $\alpha '$ terms in
the effective action of string theory. It has also been realized that
they lead to corrections to the black hole entropy. Furthermore, they
can give higher order results in the AdS/CFT correspondence. In this
talk, we will also consider them as counterterms for UV divergences of
quantum loops.

In Sect. \ref{ss:generalth}, we will review what we know about general
sugra (supergravity) and susy (supersymmetry) theories. Our preferred
method to obtain such theories uses the superconformal method, which we
review in Sect. \ref{ss:scmethod}. We will also discuss there in which
sugra theories these can be used. Then, in Sect. \ref{ss:higherderloop}
we will turn to higher-derivative sugra actions and explain the relation
with sugra loop results. We will see that we miss a lot of insight in
the possibilities for higher-derivative actions. In view of this, we
studied Dirac--Born--Infeld actions for vector multiplets, obtaining
closed expressions and exhibiting extra Volkov--Akulov type
supersymmetries. They are examples  of all order higher-derivative susy
actions. They are the deformation of the well-known lowest order
supersymmetry action, and can be considered also perturbatively in a
bottom-up construction. We will summarize this result in Sect.
\ref{ss:DBIVA}, before giving conclusions in Sect. \ref{ss:concl}.

\section{General sugra/susy theories}
 \label{ss:generalth}
An overview of possible actions with supersymmetry and supergravity has
been given in chapter 12 of the book \cite{Freedman:2012zz}, starting
from the basics. The theories considered there are `ordinary'
supersymmetry and supergravity theories, which means that the bosonic
terms in the action are at most quadratic in spacetime derivatives,
while the terms with fermions are at most linear in spacetime
derivatives. In 4 dimensions they typically contain the frame field
$e_\mu ^a$, gauge fields $A_\mu ^A$, with field strengths $F^A_{\mu \nu
}$, scalars $\varphi ^u$, gravitinos $\psi _\mu^i$, and spin-1/2
fermions $\lambda ^m$ and a Lagrangian of the form
\begin{eqnarray}
 e^{-1}{\cal L} & = & \ft12R  +\ft14(\Red{\Im {{\cal N}_{AB}}})\Blue{F^A_{\mu \nu }F^{\mu \nu
   B}}-\ft18(\Red{\Re {{\cal N}_{AB}}})e^{-1}\varepsilon ^{\mu \nu \rho \sigma }
   \Blue{F^A_{\mu \nu }F^B_{\rho \sigma }} \nonumber\\
   &&-\ft12\Red{g_{uv}}\Blue{D_\mu} \varphi ^u\Blue{D^\mu }\varphi ^v - \Blue{V}(\varphi )\nonumber\\
   &&\left\{ -\ft12\bar \psi _{\mu i}\gamma ^{\mu \nu \rho }\Blue{D_\nu} \psi _\rho
   {}^i-\ft12\Red{g_A{}^B}\bar \lambda ^A\slashed{\Blue{D}}\lambda
   _B+ \textrm{h.c.}\right\} +\ldots\,,
\end{eqnarray}
where ${\cal N}_{AB}$, $g_{uv}$ and $g_A{}^B$ are functions of the
scalars $\varphi $. In general, the possibilities for susy theories
depend on the properties of irreducible spinors in each dimension. For
theories with Minkowski signature, these can be summarised in
Table~\ref{tbl:spinors}.
\begintable{Irreducible spinors, number of components and symmetry
  properties.}{tbl:spinors}{
  \begin{tabular}{clc}
 \hline\noalign{\smallskip}
 dim & spinor & min \# components  \\
\noalign{\smallskip}\hline\noalign{\smallskip}
 2 & MW & 1\\
 3 & M & 2 \\
 4 & M & 4 \\
 5 & S & 8 \\
 6 & SW & 8\\
 7 & S & 16 \\
 8 & M & 16 \\
 9 & M & 16 \\
 10 & MW & 16\\
 11 & M & 32  \\
\noalign{\smallskip}\hline
\end{tabular}
 }
For each spacetime dimension it is indicated whether Majorana (M),
Majorana--Weyl (MW), symplectic (S) or symplectic Weyl (SW) spinors can
be defined as the `minimal spinor', and the number of real components of
this minimal spinor is given. To make a complete list, we further use
the information of what is the maximal number of susy generators in such
theories. This is based on an analysis of representations of susy in 4
dimensions, which leads to maximal $\cn=8$ for sugra, and maximal
$\cn=4$ for susy. This thus translates to maximal 32 real generators for
sugra and 16 for susy. This information is based on an analysis of
particle states i.e. states with momentum, spin and
helicity $|p^\mu,s,h\rangle$. 
One needs that susy generators transform a boson state to a fermion
state and that they square to translations, which is an invertible
operator. Considering these operators as acting from bosonic states to
fermionic states or the inverse, leads to the conclusion that there are
an equal number of bosonic and fermionic states (number of degrees of
freedom), and to the possible particle representations
\cite{Strathdee:1987jr}. The information of the maximal number of susy
generators can also be used in dimensions higher than 4, since any
higher-dimensional theory can be reduced on tori to $D=4$, keeping the
same number of susy generators. We recalled the essential elements of
the proofs here, in order to distinguish supersymmetries of this kind,
to the Volkov--Akulov supersymmetries. The latter do not transform
between such bosonic and fermionic states and should thus not be
included in the relevant counting of the number of supersymmetry
generators. Using this information leads to Table \ref{tbl:mapsusy}.
\begintable{Supersymmetry and supergravity theories in dimensions 4 to
  11.}{tbl:mapsusy}
{\tabcolsep 5pt
  $\begin{array}{| *{11}{c|} }
 D & \mbox{SUSY} & \multicolumn{2}{c|}{32} & 24 & 20 & \multicolumn{2}{c|}{16}  & 12 & 8 & 4  \\
\hline
11  & \mathrm{M} & M &   & &  & \multicolumn{2}{c|}{ }  &  &  &  \\
10  & \mathrm{MW} & \mathrm{IIA} & \mathrm{IIB} & &  &
\begin{array}{c}
  I \\
\phantom{{\cal N}}\heartsuit \\
\end{array} &  &  &  &  \\
9  & \mathrm{M} & \multicolumn{2}{c|}{{\cal N}=2 } & &  &
\begin{array}{c}
  {\cal N}=1 \\
\phantom{{\cal N}}\heartsuit \\
\end{array} &  &  &  &  \\
8  & \mathrm{M} &  \multicolumn{2}{c|}{{\cal N}=2 }&  &  &
\begin{array}{c}
  {\cal N}=1 \\
\phantom{{\cal N}}\heartsuit \\
\end{array}&  &  &  &  \\
7  & \mathrm{S} &  \multicolumn{2}{c|}{{\cal N}=4 } & &  &
\begin{array}{c}
  {\cal N}=2 \\
\phantom{{\cal N}}\heartsuit \\
\end{array}& &  &  &  \\
6  & \mathrm{SW} & \multicolumn{2}{c|}{(2,2)} &(2,1) & &
\begin{array}{c}
  (1,1) \\
\phantom{{\cal N}}\heartsuit \\
\end{array} &\begin{array}{c} (2,0)\\
\phantom{{\cal N}}\diamondsuit \\
\end{array}  &  &\begin{array}{c} (1,0) \\
\heartsuit,\diamondsuit,\clubsuit \\
\end{array}   &  \\
5  & \mathrm{S} &  \multicolumn{2}{c|}{{\cal N}=8 }  &{\cal N}=6  & & \multicolumn{2}{c|}{
\begin{array}{c}
  {\cal N}=4 \\
\phantom{{\cal N}}\heartsuit \\
\end{array}}&  &\begin{array}{c}  {\cal N}=2 \\
\phantom{{\cal N}}\heartsuit,\clubsuit \\
\end{array}  &  \\
4  & M &  \multicolumn{2}{c|}{{\cal N}=8 }  & {\cal N}=6 & {\cal N}=5 &
\multicolumn{2}{c|}{
\begin{array}{c}
  {\cal N}=4 \\
\phantom{{\cal N}}\heartsuit \\
\end{array} }  &\begin{array}{c}{\cal N}=3 \\
\phantom{{\cal N}}\heartsuit \\
\end{array}  &\begin{array}{c}  {\cal N}=2 \\
\heartsuit,\clubsuit \\
\end{array}  & \begin{array}{c}  {\cal N}=1 \\
\heartsuit,\clubsuit \\
\end{array} \\
\hline \multicolumn{2}{|c|}{ }   & \multicolumn{4}{c|}{\mathrm{SG}}  &
 \multicolumn{2}{c|}{\mathrm{SG/SUSY}} & \mathrm{SG} & \multicolumn{2}{c|}{\mathrm{SG/SUSY}}  \\
\end{array}$
 }
  An entry in the table represents the possibility to have supergravity
theories in a specific dimension $D$ with the number of (real)
supersymmetries indicated in the top row. We first repeat for every
dimension the type of spinors that can be used. Theories with up to 16
(real) supersymmetry generators allow `matter' multiplets. Considering
the on-shell states of the free theories we distinguish different kinds
of such multiplets. Those that contain a gauge field $A_\mu $ are called
vector multiplets or gauge multiplets, and are indicated in Table
\ref{tbl:mapsusy} with $\heartsuit$. Tensor multiplets in $D=\hbox{6}$
contain an antisymmetric tensor $T_{\mu \nu }$, are are indicated by
$\diamondsuit$. Multiplets with only scalars and spin-$\hbox{1/2}$
fields are indicated with $\clubsuit$. They are the hypermultiplets in
case of 8 supersymmetry generators, or the Wess--Zumino chiral
multiplets for $\cn=1$, $D=4$. At the bottom is indicated whether these
theories exist only in supergravity (SG), or also with just global
supersymmetry (SUSY).\footnote{Some exotic possibilities, like (4,0),
(2,1) theories, for which no full action exists, are omitted here.}

For each entry in the Table there are basic supergravities and
`deformations'. Basic supergravities have only gauged supersymmetry and
general coordinate transformations (and U(1)'s of vector fields). There
is no potential for the scalars, and there are only Minkowski vacua. A
deformation means that, without changing the kinetic terms of the
fields, the couplings are changed. Many deformations are 'gauged
supergravities'. That means that a Yang--Mills group is gauged,
introducing a potential. Such supergravities are produced by fluxes on
branes in string theory. There are also other deformations (e.g. massive
deformations, the superpotential in $\cn=1$ supersymmetry, \ldots ).

The embedding tensor formalism offers a way to classify the gauged
supergravities. It defines the gauge group as a subgroup of the isometry
group G, as can be seen from the covariant derivative
$\left(\partial_{\mu}-A_{\mu}{}^{M}\Red{\Theta_{M}{}^{\alpha}}
\delta_{\alpha}\right)\phi $. Here, $\alpha $ labels all the rigid
symmetries, while $M$ labels those that are gauged. The `embedding
tensor' $\Theta_{M}{}^{\alpha}$ determines which symmetries are gauged
and in which amount they contribute. E.g. the coupling constants are
part of this tensor. The tensor should satisfy a number of constraints,
whose solutions determine the possible gaugings
\cite{Cordaro:1998tx,Nicolai:2001sv,deWit:2005ub}. This thus allows to
get a complete picture of supergravities with at most two spacetime
derivatives in Lagrangian, though it  still needs more work to get all
the explicit solutions of the constraints.

For higher-derivative actions there is no such systematic knowledge.
There are various constructions of higher derivative terms, e.g. using
supersymmetric Dirac--Born--Infeld actions, but there is no systematic
construction or classification of possibilities; certainly not for
supergravity, but even not for supersymmetry.

\section{The superconformal method}
 \label{ss:scmethod}

There are various ways to construct supergravity actions. A basic way is
the order-by-order Noether method: starting from a globally symmetric
action, next order terms in the gravitational coupling constant are added
using the concepts of Noether currents. This is in fact the only
possibility for the theories with more than 16 susy generators. The
superspace method is very useful for rigid $\cn=1$ supersymmetry. However,
it becomes very difficult for supergravity. One needs many fields and many
gauge transformations to get to a supergravity action. There is also the
(super)group manifold approach, where optimal use is made of the
symmetries using constraints on the curvatures. We adhere to the method of
superconformal tensor calculus whenever possible. This method has the
advantage that it uses the nice features of superspace, like the the
structure of multiplets, but it avoids its immense number of unphysical
degrees of freedom. The extra symmetries that are used in this method
often lead to insight in the structure of a supergravity theory.

Superconformal symmetry is  the maximal extension of spacetime symmetries according to the Coleman--Mandula
theorem. What we have in mind, is not the construction of the supersymmetric completion of  Weyl
gravity, $ \int \rmd^4x\sqrt{g}\left[  R_{\mu \nu \rho \sigma }^2 - 2 R_{\mu \nu }^2 +
\ft13R^2\right]$, but the construction of Poincar{\'e} gravity,
\begin{equation}
  S_{\rm Poinc}=\int \rmd^4x\,\frac 1{2\kappa ^2}\sqrt{g}\,R\,,
 \label{SPoinc}
\end{equation}
using conformal methods, where the dimensionful gravitational coupling
constant $\kappa $ signals a breaking of the conformal symmetry. Thus,
we use the conformal symmetry as a tool for the construction of actions.
It allows us to use multiplet calculus similar to superspace, and it
makes hidden symmetries explicit.

We first explain the strategy for the construction of pure gravity in a
conformal way. One starts with a conformal coupling of a scalar field,
which will act as `compensator':
\begin{equation}
{\cal L}= -\ft12\sqrt{g}\,\phi\, \Box^C\phi=-\ft12\sqrt{g}\,\phi \Box\phi +\ft1{12}\sqrt{g}\,R\phi ^2  \,.
 \label{confactscalar}
\end{equation}
This action has local scale transformations $\delta \phi (x)=\lambda _{\rm
D}(x)\phi (x)$. These can be gauge-fixed by choosing a value
\begin{equation}
  \phi =\sqrt{6}/\kappa\,.
 \label{gaugefixedphikappa}
\end{equation}
This introduces the scale $\kappa $, indicating the breaking of
conformal symmetry. Using (\ref{gaugefixedphikappa}) in
(\ref{confactscalar}) leads to (\ref{SPoinc}). The mechanism thus starts
with a conformal invariant action, and has a Poincar{\'e} invariant action
as a result after gauge fixing. This is systematically indicated in
Table \ref{tbl:confPoinc}.
\begintable{Conformal construction of Poincar{\'e} gravity}{tbl:confPoinc}
  {    \begin{tabular}{c}
 \fbox{Conformal gauge multiplet coupled to a scalar} \phantom{heelveelzever}\ \ {\small local conformal }\\
   \Big \Downarrow \quad gauge fix non-conformal symmetries \\
  \phantom{conformal}\fbox{Poincar{\'e} gravity}\hfill {\small local Poincar{\'e} symmetry}
\end{tabular}
    }

For the supersymmetric theories, a similar construction allows to get
more insight in the structure of supergravity actions. A main difference
between supersymmetry and supergravity is that multiplets have a clear
structure in supersymmetry, but after coupling to supergravity they
often get mixed, and they are not clearly identifiable in the final
action. In another language: superfields are an easy conceptual tool for
globally supersymmetric theories. With the similar method as described
above for gravity, supergravity can also be obtained by starting with an
action with superconformal symmetry and gauge fixing the superfluous
symmetries. This is especially useful for matter-coupled supergravities.
Before the gauge fixing, everything looks like in global supersymmetry,
just adding covariantizations for the superconformal symmetries. Only
after the gauge fixing, the multiplets get mixed.

To elucidate the superconformal symmetry, it is useful to consider it in
the way of transformations of supermatrices of the form
\begin{equation}
  \pmatrix{
  \mbox{conformal algebra}&Q,S\cr Q,S&R\mbox{-symmetry}
  }
  \,.
\label{superconformalmatrix}
\end{equation}
$Q$ is the
ordinary supersymmetry and $S$ is the extra, `special' supersymmetry.
The $R$-symmetry depends on the dimension and extension of
supersymmetry.
It is clarifying to order the generators according to their
weight under dilatations (here for the $\cn=1$ superconformal algebra)
\begin{eqnarray}
 1 & : & P_a  \nonumber\\
 \ft12 & : & Q \nonumber\\
 0 & : & D\,,\ M_{ab }\,,\ T\nonumber\\
 -\ft12 & : & S \nonumber\\
 -1 & : & K_a \,.
\label{weightsSCAlg}
\end{eqnarray}
$P_a$, $D$, $M_{ab}$ and $K_a$ are the conformal generators. The
$R$-symmetry is in this case just U(1), whose generator is indicated by
$T$. The weights in the first column of (\ref{weightsSCAlg}) determine the commutators involving $D$, for example
\begin{equation}
  [D,Q]=\ft12 Q\,, \qquad [D,S]=-\ft12 S\,.
\label{DQcomm}
\end{equation}
As we discussed above, $T$ is an $R$-symmetry. All (anti)commutators are
consistent with the weights, e.g.
\begin{eqnarray}
   &   & \left\{ Q_\alpha ,Q^\beta \right\} =
   -\ft12 (\gamma^a)_\alpha{}^\beta P_a\, ,
 \qquad
 \left\{ S_\alpha ,S^\beta \right\} =
   -\ft12 (\gamma ^a)_\alpha{}^\beta K_a \,, \nonumber\\
   &   &  \left\{ Q_\alpha ,S^\beta \right\} =
   -\ft12  \delta_\alpha{}^\beta D
   -\ft14 (\gamma^{ab})_\alpha{}^\beta M_{ab}
   +\ft12\rmi(\gamma_*)_\alpha{}^\beta T\,.
   \label{QSalgebra}
\end{eqnarray}

The strategy for the superconformal construction of $\cn=1$ supergravity
is analogous as for gravity in Table~\ref{tbl:confPoinc}. It is depicted
in Table \ref{tbl:strategyN1}
\begintable{Superconformal construction of pure $\cn=1$ supergravity}
{tbl:strategyN1}    {    \begin{tabular}{l}
\phantom{heelveelzever} \fbox{\begin{minipage}[l]{8cm}Superconformal gauge multiplet ($\cn =1$ Weyl multiplet) \\ coupled to
a
  chiral multiplet\end{minipage}} \\[2mm]
  \phantom{superChiral multiplet} \quad
  \Big \Downarrow \quad \begin{minipage}[l]{8cm}
  \small gauge fix dilatations, \\ special conformal transformations, \\
  local U(1)-symmetry and special supersymmetry \end{minipage}
  \\[4mm]
  \phantom{superconformal gau}\fbox{pure $\cn=1$ supergravity}
\end{tabular}}


A similar scheme holds for $\cn=4$ supergravity  \cite{Bergshoeff:1981is,deRoo:1984gd} as shown in Table
\ref{tbl:strategyN4}.
\begintable{Superconformal construction of pure $\cn=4$ supergravity}
{tbl:strategyN4}    {    \begin{tabular}{l}
\phantom{heelveelzever} \fbox{\begin{minipage}[l]{8cm}Superconformal gauge multiplet ($\cn =4$ Weyl multiplet) \\ coupled to
6 gauge compensating multiplets (on-shell)\end{minipage}} \\[2mm]
  \phantom{superChiral multiplet} \quad
  \Big \Downarrow \quad \begin{minipage}[l]{8cm}
  \small gauge fix dilatations, \\ special conformal transformations, \\
  local SU(4), local U(1) and special supersymmetry \end{minipage}
  \\[4mm]
  \phantom{superconformal gau}\fbox{pure $\cn=4$ Cremmer-Scherk-Ferrara supergravity}
\end{tabular}}
The special feature is that the gauge compensating multiplets are
on-shell multiplets. Remember that in any case the action should be
invariant without use of the field equations, but the algebra of the
symmetries may close only modulo field equations. However, the problem
is that in this way there is no flexibility in the field equations. They
are already fixed by the supersymmetry transformation laws. This gives
thus a problem when we want to modify the action with higher-derivative
terms, since then the field equations will change. Therefore,
higher-derivative terms cannot be added to $\cn=4$ supergravity without
a modification of the field equations. The hypermultiplets of $\cn=2$
supergravity already have this feature of an `on-shell algebra' (at
least for a generic hyper-K{\"a}hler manifold). The $\cn=4$ gauge multiplets
also share this property. This is especially relevant since they are
compensating multiplets. It implies that the supersymmetry
transformations of the $\cn=4$ super-Poincar{\'e} theory can only close
modulo field equations. But one can apply the superconformal method.

In which supergravity theories can we use the superconformal methods~?
There are two necessary ingredients. First, one should have a
superconformal algebra. Second, there should be compensating multiplets.
Which theories allow superconformal algebras was already analysed by W.
Nahm \cite{Nahm:1978tg}. He analysed in which simple superalgebras the
conformal algebra $\soa(D,2)$ is a factor in the bosonic subalgebra.
This lead to Table \ref{tbl:scalgebras} (also a long list of
superconformal algebras exist for $D=2$).
\begintable{Superconformal algebras}{tbl:scalgebras}
{\begin{tabular}{llllr}\hline
$D$& supergroup&conf & $R$ & ferm.\\ \hline
$3$&$ \OSp(N|4)   $&$\SO(3,2)=\Sp(4)\ $&$  \SO(N) $&$ 4N $\\
$4$&$ \SU(2,2|N)  $&$\SO(4,2)=\SU(2,2) $&$ \U(N)$ for $N\neq 4$&$ 8N$\\
   &              & &$   \SU(4)$ for $N= 4$& \\
$5$& $F^2(4) $  &$\SO(5,2) $&   $  \SU(2) $ & 16 \\
$6$&$ \OSp(8^*|2N) $&$\SO(6,2)=\SO^*(8)$ &$ \USp(2N)$& $16N$ \\
\hline
\end{tabular} }
In each case the bosonic subgroup contains  the covering
group\footnote{The equality sign in the `conf' column of this Table is
only valid at the level of the algebra.} of $\SO(D,2)$, such that spinor
representations are possible, and a compact $R$-symmetry group. The last
column gives the number of real supersymmetry generators. Other
superconformal algebras have been considered where the conformal algebra
is not a factor, but still a subalgebra of the bosonic part of the
superalgebra. E.g. $\SO(11,2)\subset \Sp(64)\subset \OSp(1|64)$
\cite{vanHolten:1982mx,D'Auria:2000ec}. However, these have not been
successfully applied for constructing actions. Thus, the superconformal
methods are restricted to the dimensions and extensions that appear in
Table \ref{tbl:scalgebras} and furthermore to a number of supersymmetries
$\leq 16$, such that compensating multiplets exist.\footnote{For $D=10$
with 16 supersymmetries, a superconformal formulation, not based on a Lie
superalgebra but rather on a soft algebra has been found in
\cite{Bergshoeff:1983az}.} This leads to those indicated in boxes in Table
\ref{tbl:DNsc}.
\begintable{Supergravity theories for which superconformal methods can be used}{tbl:DNsc}
{\tabcolsep 5pt
  $\begin{array}{| *{11}{c|} }
 D & \mbox{SUSY} & \multicolumn{2}{c|}{32} & 24 & 20 & \multicolumn{2}{c|}{16}  & 12 & 8 & 4  \\
\hline
11  & \mathrm{M} & M &   & &  & \multicolumn{2}{c|}{ }  &  &  &  \\
10  & \mathrm{MW} & \mathrm{IIA} & \mathrm{IIB} & &  &
I &  &  &  &  \\
9  & \mathrm{M} & \multicolumn{2}{c|}{{\cal N}=2 } & &  &
  {\cal N}=1
 &  &  &  &  \\
8  & M &  \multicolumn{2}{c|}{{\cal N}=2 }&  &  &
  {\cal N}=1
&  &  &  &  \\
7  & S &  \multicolumn{2}{c|}{{\cal N}=4 } & &  &
  {\cal N}=2
& &  &  &  \\
6  & SW & \multicolumn{2}{c|}{(2,2)} &(2,1) & &
  (1,1)
 &\ 
\fbox{(2,0)}\ &  & \ \fbox{(1,0)}\   &  \\
5  & S &  \multicolumn{2}{c|}{{\cal N}=8 }  &{\cal N}=6  & &
\multicolumn{2}{c|}{
  {\cal N}=4 }&  &\ \fbox{${\cal N}=2$}\    &  \\
4  & M &  \multicolumn{2}{c|}{{\cal N}=8 }  & {\cal N}=6 & {\cal N}=5 &
\multicolumn{2}{c|}{
\ \fbox{${\cal N}=4$}\ } &\ \fbox{${\cal N}=3$}\  &\ \fbox{${\cal N}=2$}\   & \ \fbox{${\cal N}=1$}\   \\
\hline
\end{array}$
 }

\section{Higher derivative sugra actions and sugra loop results}
 \label{ss:higherderloop}

For many years it was believed that supergravity could not be a finite
theory. However, since the calculations of \cite{Bern:2007hh} revealed
the 3-loop finiteness of $\cn=8$, $D=4$ supergravity, we realize that
quantum supergravity has more surprising features than we understood so
far. In  \cite{Bern:2009kd} the result was extended to 4 loops and even
to $D=5$. But then, also $\cn=4$ supergravity in $D=4$ turned out to be
finite up to 3 loops \cite{Bern:2012cd} (and further results followed
for $D=5$). This brings us to reflections on the nature of supergravity
and possible counterterms. Divergences would imply that supersymmetric
counterterms should exist (or there should be supersymmetric anomalies).
But our present knowledge on higher-derivative terms in supergravity is
not sufficient to be sure about which invariants can be consistently
defined.

\subsection{Superconformal methods for the $\cn=2$ example}
Superconformal methods have been used to construct higher-derivative
supergravities, starting with the work of S. Cecotti and S. Ferrara
\cite{Cecotti:1986gb}. Especially for $\cn=2$ supergravity, the tensor
calculus allows us to construct various terms \cite{deWit:2010za}. 
The constructions use tensor calculus with chiral multiplets, which are
similar to chiral superfields. The multiplets contain fields
\begin{equation}
  S=\{X,\Omega _i,\ldots ,C\}\,.
 \label{SchiralN2}
\end{equation}
Any sum and product of these gives another chiral multiplets. These
manipulations allow `tensor calculus'. A useful tool is the kinetic
multiplet of a chiral multiplet (which is also chiral) and starts with
the complex conjugate of the highest component of a chiral multiplet:
\begin{equation}
  \mathbb{T}(\bar S)=\{\bar C,\ldots \}\,.
 \label{kineticchiralN2}
\end{equation}
To construct higher-derivative terms, one needs also another chiral
multiplet, formed from the $\cn=2$ Weyl multiplet
\begin{equation}
  W^2=\{T_{ab}^-T^{ab\,-},\ldots \}\,.
 \label{W2chiralN2}
\end{equation}
It starts from the square of an auxiliary field (antisymmetric tensor)
of the Weyl multiplet. One can then use tensor calculus on these
multiplets to construct new chiral multiplets, of which the highest
components defines actions. In order to be able to define these in the
superconformal framework, one has to take into account the dilatation
symmetry. This implies that the function of chiral multiplets that is
used to construct actions should satisfy homogeneity properties. Using
such homogeneous functions of the chiral multiplets, one obtains
supergravity theories using superconformal covariantization of the
expressions used for global supersymmetry. Hence this leads to many
possibilities, which are invariants contributing to the entropy and
central charges of black holes.

In order to see how these actions lead to DBI theories, $R^4$ actions
are considered in \cite{Chemissany:2012pf}, using the above-mentioned
constructions with
\begin{equation}
  \left[ S^2 + \lambda \frac{W^2}{S^2}\mathbb{T}\left(\frac{\bar W^2}{\bar S^2}\right) \right] _C\,.
 \label{SWconstrDBI}
\end{equation}
It uses the action formula `$C$', which means in global supersymmetry
the highest component of th chiral multiplet. In superconformal
calculus, there are some correction terms involving the gravitino, to
obtain local conformal symmetry. $S$ is the chiral compensating
multiplet (which due to constraints is in fact a vector multiplet).
Using just the first term in (\ref{SWconstrDBI}) would lead to pure
supergravity.\footnote{In fact, a second compensating multiplet is
necessary in $\cn=2$, but we do not discuss this here, since this can be
neglected for the present purposes.} The second term in
(\ref{SWconstrDBI}) uses the multiplet (\ref{W2chiralN2}) and the
construction of a kinetic multiplet (\ref{kineticchiralN2}). The powers
of $S$ are chosen in order to satisfy the homogeneity properties leading
to conformal-invariant actions. That second term is taken with a
coupling constant $\lambda $, in which an expansion will be considered.

Apart from a term of the form $\lambda C_{\cdot \cdot \cdot \cdot }^4$,
where $C_{\cdot \cdot \cdot \cdot }$ is the Weyl tensor, and thus
creating terms of the form $R^4$, the action formula in
(\ref{SWconstrDBI}) produces also terms of the type $\lambda (\partial
T)^4$, where $T$ stands for the auxiliary field of the Weyl multiplet.
In the standard supergravity action, the field equations imply that $T$
is on-shell proportional to the graviphoton. For  the action
(\ref{SWconstrDBI}), we get, symbolically
\begin{equation}
  T_{ab}= \frac{2}{X} F_{ab} + \lambda (\partial ^4 T^3)_{ab}\,,
 \label{feTlambda}
\end{equation}
where $X$ is the scalar of the compensating multiplet, which is in the
Poincar{\'e} theory dependent on $\kappa $ similar to
(\ref{gaugefixedphikappa}). This equation is solved recursively, and we
thus get an expression with an infinite number of higher derivative
terms with higher and higher powers of the graviphoton $F$:
\begin{equation}
  T_{ab}=\frac{2}{X} F_{ab} + \lambda (\partial ^4 F^3)_{ab} + \lambda ^2\partial ^4 F^2\partial ^4F^3+\ldots \,.
 \label{Tinfinite}
\end{equation}
The action with auxiliary field eliminated leads to a DBI-type action
with higher derivatives
\begin{equation}
  S_{\rm deformed}= -\ft14 F^2 + \lambda (\partial F)^4 +\lambda ^2 \partial ^8F^6+\ldots \,.
 \label{SdeformedN2}
\end{equation}
Note that before the elimination of the auxiliary field, this action has
a finite number of terms with auxiliary fields. The infinite series is
produced by the elimination of the auxiliary fields. They lead thus to a
deformation of the lowest order action in powers of $\lambda $. At the
same time also the transformation laws are deformed. Again, the
transformation laws are finite expressions before the elimination of the
auxiliary fields. E.g. for the gravitino transformation
\begin{equation}
  \delta\psi_\mu^i = D _\mu
   \epsilon^i   -\ft1 {16} \gamma^{ab} T_{ab}^-\varepsilon^{ij}\gamma_\mu\epsilon_j -\gamma_\mu\eta^i\,,
 \label{delpsiconfN2}
\end{equation}
where the covariant derivative uses the superconformal connections, and
$S$-supersymmetry with parameter $\eta ^i$ is included. Then the
on-shell value of the auxiliary fields is used as a power series in
$\lambda $:
\begin{equation}
  \phi _{\rm aux}= \phi^{(0)} _{\rm aux}+\Delta \phi _{\rm aux}\,,\qquad \Delta \phi _{\rm aux}=\sum_{n=1}\lambda ^n\phi^{(n)} _{\rm aux}\,.
 \label{phiauxexpand}
\end{equation}
This leads, with (\ref{Tinfinite}), to deformations in the supersymmetry
transformation law of the gravitino of the form \cite{Chemissany:2012pf}
\begin{equation}
  \Delta\psi_\mu^i = -4\lambda [\partial ^4F^3]_\mu {}^\nu \gamma _\nu \epsilon ^i +\ldots \,.
 \label{DeltapsiF4}
\end{equation}
Here also contributions have been used that originate from the
`decomposition law' expressing the parameter $\eta ^i$ in terms of
$\epsilon ^i$ after gauge fixing of $S$-supersymmetry.

We conclude that the tensor calculus allows us to obtain
higher-derivative terms, determined first off-shell, which can lead to
deformations of the action and transformation laws on-shell. They are
obtained from (broken) superconformal actions. For pure gravity, the
3-loop counterterm that contains $R^4$ is obtained from the local
conformal expression
\begin{equation}
  \int \rmd^4\sqrt{g}\,\phi ^{-4}(C_{\mu \nu \rho \sigma }C^{\mu \nu \rho \sigma})^2\,,
 \label{N0ct}
\end{equation}
where $\phi $ is the compensating scalar and $C_{\mu \nu \rho \sigma }$
is the Weyl tensor. For $\cn=2$ a superconformal $R^4$ counterterm can
be obtained from the $\lambda $-term in (\ref{SWconstrDBI}). What do we
know about $\cn=4$, where miraculous cancelations have been found?

\subsection{Problem and conjecture for $\cn=4$ supergravity}

The problem is that it is not easy to construct counterterms for $\cn=4$
supergravity. We cannot multiply the compensating multiplets to suitable
powers, and thus we cannot make constructions as those for $\cn=2$. The
essential problem is that the algebra of supersymmetry holds only on
shell. When we would like to write a modified action, then it implies
modified field equations, and thus the transformations have to be
modified (or in other words: the structure of the multiplets). For
$\cn=2$, deformed transformations could be found due to the possibility
to work first with auxiliary fields. The field equations for the latter
lead to deformed transformation laws on shell. For $\cn=4$ we do not
know auxiliary fields. How can we then establish the the existence or
non-existence of the consistent order by order deformation of $\cn=4$
supergravity?

This question lead to the conjecture made in \cite{Ferrara:2012ui}. If
such counterterms do not exist, this may explain finiteness results (if
meanwhile the explicit calculations do not find that $\cn=4$, $D=4$ is
divergent at higher loops). Until invariant counterterms are constructed
we have no reason to expect UV divergences. We can also conjecture that
such counterterms should be broken superconformal expressions, if
conformal symmetry is more than a classical symmetry. Thus there are two
points of view. The first one is that legitimate counterterms are not
available yet, and we still have to construct them. The second one is
that legitimate counterterms are not available, and cannot be
constructed, offering an explanation of finiteness.

In fact, if the UV finiteness will persist in higher loops, one would like
to view this as an opportunity to test some new ideas about gravity. One
possible idea is that superconformal symmetry, used in the classical
theory as a tool to construct actions, is more fundamental and has also a
quantum significance. As mentioned in Sect. \ref{ss:scmethod}, the
classical theory can be obtained from gauge fixing a superconformal
action. In that way, the Planck mass appears only in the gauge-fixing
procedure. This looks analogous to the appearance of the masses of $W$ and
$Z$ vector mesons in the standard model. They are not present in the
gauge-invariant action, and show up when the gauge symmetry is
spontaneously broken. In the unitary gauge these masses give the
impression of being fundamental. In the renormalizable gauge, where the UV
properties are analysed, they are absent. One may hope that a similar
understanding can be obtained in the future to give a more fundamental
significance to the superconformal symmetry. The possible non-existence of
(broken) superconformal-invariant counterterms and anomalies in $\cn=4$,
$D=4$ supergravity could then explain the 'miraculous' results of the
quantum calculations.

Such ideas would give a simple explanation of the 3-loop finiteness and
predict perturbative UV finiteness in higher loops. The same conjecture
applies to higher-derivative superconformal invariants and to the
existence of a consistent superconformal anomaly. Also for the latter,
one may either say that we still have to understand how to construct
such an anomaly, or maybe it does not exist. Therefore, the conjecture
is economical, sparing in the use of resources: either the local $\cn=4$
superconformal symmetry is a good symmetry, or it is not. The conjecture
is falsifiable by the $\cn=4$ 4-loop computations (which are already
underway, as we heard during the conference). If the conjecture survives
these computations (if they show further UV finiteness), then this gives
a further hint that the models with superconformal symmetry serve as a
basis for constructing a consistent quantum theory where the Planck mass
appears only in the process of gauge fixing the superconformal symmetry.
However, it is also falsifiable by our own calculations: if we find a
way to construct (non-perturbative) superconformal invariants that can
serve as counterterms, then this conjecture is circumvented. We will
start to search in that direction, following a quote of R. Feynman: ``We
are trying to prove ourselves wrong as quickly as possible, because only
in that way can we find progress.''

\section{Dirac--Born--Infeld - Volkov--Akulov and deformation of supersymmetry}
 \label{ss:DBIVA}

The main problem for the superconformal construction of counterterms in
$\cn=4$ supergravity is thus that the compensating multiplets have only
been defined with transformations that close on-shell using the field
equations of the 2-derivative action. These compensating multiplets are
vector multiplets. In our recent work \cite{Bergshoeff:2013pia} we
search for deformations of vector multiplet actions such that
higher-derivative terms occur. We will find all-order higher derivative
globally supersymmetric invariant actions. They are of the
Dirac--Born--Infeld (DBI) type, and have extra symmetries, of
Volkov--Akulov (VA) type. The latter are not yet $S$-supersymmetry
transformations that we would like in the context of the superconformal
programme mentioned above, but we will comment on this at the end.

We will consider vector multiplets with a gauge vector and a spinor
field. We want that the supersymmetry algebra is closed, but not
necessary off-shell, since the main problems that we want to address are
theories with only an on-shell closed algebra. A gauge vector in $D$
dimensions has $D-2$ on-shell degrees of freedom,\footnote{All these
ingredients are well defined and discussed in \cite{Freedman:2012zz}.}
while a spin-1/2 fermion has on-shell half the number of degrees of
freedom of the number
of components of the spinor. Considering 
Table \ref{tbl:spinors} shows that one can have an equal number of
bosonic and fermionic degrees of freedom for these fields in the cases
$D=10$ with Majorana--Weyl spinors, $D=6$ with symplectic Majorana--Weyl
spinors, $D=4$ with Majorana spinors, and even $D=3$ with Majorana
spinors. Comparing with Table \ref{tbl:mapsusy} shows that these are the
maximal dimensions to have vector multiplets for supersymmetries with
16, 8, 4 and 2 generators. Other vector multiplets are obtained from
these by dimensional reduction, which generates also scalar on-shell
degrees of freedom.

These theories are described by an action of the form\footnote{With
respect to \cite{Bergshoeff:2013pia} all barred spinors are multiplied
with a factor $-1/2$ in order to agree with the normalizations as in
(\ref{QSalgebra}) and \cite{Freedman:2012zz}.}
\begin{equation}
  S = \int \rmd^{D}x\,\big \{-\ft{1}{4} (F_{\mu\nu})^2 -\ft12 \bar\lambda
\slashed{\partial} \lambda \big\}\,.
 \label{SYM}
\end{equation}
They are invariant under supersymmetry transformations\footnote{We use
here $\Gamma $ rather than $\gamma $ for the gamma matrices in the $D$,
to distinguish them later from the 4-dimensional matrices, see
(\ref{Gamma10gamma4}).}
\begin{equation}
  \delta_\epsilon  A_\mu = -\ft12{\bar\epsilon}\Gamma_\mu\lambda \,,\qquad
\delta_\epsilon \lambda = \ft{1}{4}\Gamma^{\mu \nu}F_{\mu \nu}\epsilon\,,
 \label{susySYM}
\end{equation}
where the spinors are of the appropriate type mentioned before, and for
the case of symplectic Majorana-Weyl spinors also the extension index
$i=1,2$ has been suppressed with the understanding that e.g.
\begin{equation}
  {\bar\epsilon}\Gamma_\mu\lambda ={\bar\epsilon}^i\Gamma_\mu\lambda_i=\varepsilon ^{ij}{\bar\epsilon}_j\Gamma_\mu\lambda_i\,.
 \label{ijsympl}
\end{equation}
The action has also an extra trivial (global) fermionic shift symmetry
\begin{equation}
  \delta_\eta  A_\mu = 0\,,\qquad  \delta_\eta \lambda= -\ft{1}{2\alpha }\eta\,,
 \label{trivialsusyYM}
\end{equation}
where the normalization with a constant $\alpha $ has been used in order
to match with formulas that will follow below.
\subsection{The bottom-up approach}
We first attempt a `bottom-up' approach. This means that we define a
deformation of the action with terms proportional to a parameter
$\alpha$, and adapt simultaneously the transformation laws. In this we
follow \cite{Bergshoeff:1986jm}, where this was considered for $D=6$,
and an action was obtained of the form
\begin{eqnarray}
S &=& \int \rmd^{D}x\,\big\{-\ft{1}{4} F^2 -\ft12 \bar\lambda \slashed{\partial} \lambda\big\} +\alpha c_4 F^{\mu\nu}\bar\lambda\Gamma_\mu\partial_\nu\lambda \nonumber\\[.2truecm]
&&+\ft{1}{8}\alpha^2\Big[\Tr F^4 -\ft{1}{4}\, \left(F^2\right)^2 -2 (1+4c_4^2) \big(F^2\big)^{\mu\nu}\bar\lambda\Gamma_\mu\partial_\nu\lambda\nonumber\\[.2truecm]
&&-\ft12 (1-4c_4^2)F_\mu{}^\lambda\big(\partial_\lambda F_{\nu\rho}\big)\bar\lambda\Gamma^{\mu\nu\rho}\lambda -\ft14(c_1+8c_4^2) F^2\bar\lambda \slashed{\partial}\lambda\nonumber\\[.2truecm]
&&+\ft{1}{4}c_2F_{\mu\nu}\big(\partial_\lambda F^\lambda{}_\rho\big)\bar\lambda \Gamma^{\mu\nu\rho}\lambda+\ft{1}{4} (c_3+4c_4^2)F_{\mu\nu}F_{\rho\sigma}\bar\lambda
\Gamma^{\mu\nu\rho\sigma} \slashed{\partial}\lambda \Big ]\nonumber\\
&&+{\cal O}(\alpha ^2\lambda ^4)+{\cal O}(\alpha ^3)\,.
\label{Sbottomup}
\end{eqnarray}
The parameters $\lambda _i$ are undetermined. However, they are all
related to field redefinitions
\begin{eqnarray}
A_\mu(0) &=& A_\mu + \ft{1}{32}\alpha^2 c_2 F^{\nu\rho}\bar\lambda\Gamma_{\mu\nu\rho}\lambda\,,\nonumber\\
\lambda(0) &=& \lambda + \ft{1}{2}\alpha c_4 F_{\mu\nu}\Gamma^{\mu\nu}\lambda +\ft{1}{32}\alpha^2c_1 F^2 \lambda
 -\ft{1}{32}\alpha^2c_3F_{\mu\nu}F_{\rho\sigma}\Gamma^{\mu\nu\rho\sigma}\lambda\,,
\end{eqnarray}
where on the right-hand side are the fields corresponding to $c_i=0$,
and on the left-hand side those for arbitrary $c_i$. Hence, up to these
redefinitions, the answer is unique up to this order. Remark e.g. that
it contains in the bosonic part the unique combination
\begin{equation}
\Tr F^4 -\ft{1}{4}\, \left(F^2\right)^2\,,\qquad \Tr F^4\equiv  F_{\mu \nu }F^{\nu
\rho }F_{\rho \sigma }F^{\sigma \mu }\,,\qquad F^2= F_{\mu \nu
}F^{\mu \nu }\,.
 \label{F4combination}
\end{equation}
Also the transformation laws are deformed with respect to
(\ref{susySYM}). As well ordinary supersymmetry transformations
(parameter $\epsilon $) as the extra supersymmetry (\ref{trivialsusyYM})
can be defined. E.g. for the latter we have now
\begin{eqnarray}
\delta_\eta  A^\mu&=&-\ft{\alpha}{8 }\bar \eta F^{\nu\mu }\Gamma_\nu \lambda-\ft{\alpha}{16 }\bar \eta \Gamma^{\mu\nu\rho} F_{\nu\rho }\lambda
 +\ft{1}{32}\alpha c_2 F_{\nu\rho}\bar\eta\Gamma^{\mu\nu\rho}\lambda +{\cal O}(\alpha \eta \lambda ^3)+{\cal O}(\alpha ^2)
\,,\nonumber\\
  \delta_\eta \lambda&=& -\ft{1}{2\alpha }\eta+\alpha\left[-\ft{1}{32} F^2-\ft{1}{64}\Gamma^{\mu\nu\rho\sigma} F_{\mu\nu}F_{\rho\sigma} \right] \eta\nonumber\\
&&+\ft{1}{4}c_4 F_{\mu\nu}(c)\Gamma^{\mu\nu}
\left[ \eta -\ft{1}{2}\alpha c_4 F_{\rho \sigma }(c)\Gamma^{\rho \sigma }\eta \right]
\nonumber\\
&&+\ft{1}{64}\alpha c_1 F^2 \eta
 -\ft{1}{64}\alpha c_3F_{\mu\nu}F_{\rho\sigma}\Gamma^{\mu\nu\rho\sigma}\eta+{\cal O}(\alpha \eta \lambda ^2)+{\cal O}(\alpha ^2)\,.
\end{eqnarray}
It turns out that we can write this for all $D=10,6,4,3$ with the
appropriate spinors types (Majorana, Majorana--Weyl, symplectic
Majorana--Weyl) as mentioned above . The only spinor properties that we
need are the Majorana flip relations, like
\begin{equation}
  \bar \lambda _1 \Gamma ^\mu \lambda _2=- \bar \lambda _2 \Gamma ^\mu  \lambda _1\,,\qquad
  \bar \lambda _1 \Gamma ^{\mu\nu \rho } \lambda _2= \bar \lambda _2 \Gamma ^{\mu\nu \rho }  \lambda _1\,,
 \label{Majoranaflip13}
\end{equation}
and the cyclic Fierz identity
\begin{equation}
  \Gamma_\mu \lambda_1\bar{\lambda}_2\Gamma^\mu \lambda_3+\Gamma_\mu \lambda_2\bar{\lambda}_3\Gamma^\mu \lambda_1+\Gamma_\mu \lambda_3\bar{\lambda}_1\Gamma^\mu \lambda_2=0\,.
 \label{cyclicFierz}
\end{equation}
These are valid for all these cases. Note that all bilinears in spinors
contain odd-rank gamma matrices, as is consistent with the fact that the
spinors are all of the same chirality in $D=10$ and $D=6$. But this
property holds also for e.g. $D=4$.

The results look very complicated and it seems hopeless to continue this
to all orders in $\alpha $ and adding higher order spinor terms.

\subsection{The top-down approach}
 \label{ss:topdown}
We  \cite{Bergshoeff:2013pia} found a solution to the problem of the
construction of the infinite series of deformations starting from the
$\kappa $-symmetric action for D$p$ branes. This action is of the form
\begin{equation}
  S_{\rm DBI} +S_{\rm WZ} =  -\frac{1}{\alpha^2}
\int \rmd^{p+1} \sigma\, \sqrt{- \det (G_{\mu\nu} + \alpha
{\cal F}_{\mu\nu})} +\frac{1}{\alpha^2}\int \Omega_{p+1} \,,
 \label{SDp}
\end{equation}
where the first term is a DBI action, and $\kappa $-supersymmetry
implies that it should be complemented with a Wess--Zumino (WZ) term in
terms of an appropriate $(p+1)$-form $\Omega _{p+1}$ (see e.g. (45) in
\cite{Aganagic:1996nn}). In the DBI term appear
\begin{eqnarray}
 &&G_{\mu\nu} \equiv  \eta_{mn} \Pi_\mu^m \Pi_\nu^n \ , \qquad \Pi_\mu^m \equiv
\partial_\mu X^m +\ft12\bar\theta \Gamma^m \partial_\mu \theta\,, \nonumber\\
   &   & {\cal F}_{\mu\nu} \equiv F_{\mu\nu}+\alpha^{-1} \bar{\theta} \sigma_3
\Gamma_{m}\partial_{[\mu}\theta\left(\partial_{\nu]} X^{m}
+\ft{1}{4} \bar{\theta}\Gamma^{m}\partial_{\nu]}\theta\right)\,.
\end{eqnarray}
We consider these actions in the context of the IIB theory, and thus
$X^m$ with $m=0,\ldots ,9$ denote the spacetime coordinates of the
$D=10$ theory. The coordinates on the brane are indicated by $\mu
=0,\ldots ,p$, and $p$ should be odd. $\theta $ is a doublet of
Majorana--Weyl spinors, of which we omit again the extension index.
$F_{\mu \nu }$ is an Abelian field strength.

This action has the following symmetries. First, there is a rigid
supersymmetry doublet parameter $\epsilon ^1$, $\epsilon ^2$. There is
also rigid Poincar{\'e} symmetry in $D=10$. Furthermore, there are local
symmetries on the brane. On the bosonic side these are the worldvolume
general coordinate transformations. Furthermore there is the $\kappa
$-supersymmetry doublet. Effectively only half of these are present,
since they this is a reducible symmetry, which means that it appears
only in the form
\begin{equation}
  \delta_\kappa\theta=(1+\Gamma)\kappa\,,
 \label{kappareducible}
\end{equation}
where $\Gamma $ is a matrix such that $(1+\Gamma )$ is a projection on
half of the spinor space.

Though this has been obtained from IIB superstring theory in $D=10$, it
turns out that the action (\ref{SDp}) has also the same symmetries when
we consider $D=6$, just changing the index range to $m=0,\ldots ,5$ and
using symplectic Majorana--Weyl spinors. This implies that we consider
the $(2,0)$ theory in the $D=6$, 16 supersymmetries entry of Table
\ref{tbl:mapsusy}.  This theory is often called iib. The action has then
also a brane interpretation, (using again odd $p$)
\cite{Bergshoeff:2012jb} as has been clarified in the talk of E.
Bergshoeff in this conference. Moreover, we can also consider it
solutions of $D=4$, $\cn=2$ supergravity with worldvolume action as in
(\ref{SDp}) (thus $m=0,\ldots ,3$ and $p=3$ or 1).

We then gauge-fix local symmetries imposing for a $p$-brane (describing
here the embedding in $D=10$, but the other cases are obtained by
changing the range of indices)
\begin{eqnarray}
   &   & X^m =\{\delta^{m^\prime}_{\mu}\sigma^{\mu}, \phi^I\}\,, \quad  m^\prime
=0,1,\dots,p\,,\quad I=1,\dots,9-p \nonumber\\
   &   & \theta =(\theta^{1} = 0, \, \theta^2 \equiv
\alpha\lambda)\,.
\label{gaugefixgctkappa}
\end{eqnarray}
The first line fixes the worldvolume general coordinate transformations
by identifying the coordinates in the embedding spacetime with the
worldvolume coordinates. This leaves $9-p$ scalars. In the second line,
the effective $\kappa $-symmetry is fixed, and the remaining coordinate
is renamed $\lambda $ in order to make the connection with the down-up
approach. These gauges lead to decomposition laws, implying that the
parameters of the worldvolume general coordinate transformations and
$\kappa $-symmetry become functions of the remaining (global)
symmetries. There are thus two, deformed, fermionic symmetries $\epsilon
^1$ and $\epsilon ^2$. Two combinations of these symmetries are called
$\epsilon $ and $\zeta $, and can be related to the $\epsilon $ and
$\eta$ symmetries of the bottom-up approach.

We first consider the action for the case $p=9$ in this gauge, which
reduces (\ref{SDp}) to
\begin{equation}
  S =  -\frac{1}{\alpha^2} \int \rmd^{10} x\,\left\{ \sqrt{- \det (G_{\mu\nu} + \alpha {\cal F}_{\mu\nu})}-1\right\}\,,
 \label{Sd10gf}
\end{equation}
where
\begin{eqnarray}
 &&G_{\mu\nu} = \eta_{mn} \Pi_\mu^m \Pi_\nu^n \ , \qquad
\Pi_\mu^m = \delta_\mu^m +\ft12 \alpha^2\bar\lambda \Gamma^m \partial_\mu \lambda \,, \nonumber\\
&&{\cal F}_{\mu\nu} \equiv F_{\mu\nu} -\alpha \bar \lambda \Gamma _{[\nu }\partial _{\mu ]}\lambda\,,\qquad \mu=0,1,...,9\,, \qquad m=0,1,...,9\,.
\end{eqnarray}
This action possesses 16 $\epsilon$ transformations, which are
deformations of the Maxwell supermultiplet supersymmetries:
\begin{eqnarray}
\delta_{\epsilon} \lambda &=& -\ft{1}{2\alpha}\left( \unity-
\beta\right)\epsilon
+\ft{1}{4}\alpha\partial_{\mu}\lambda\bar{\lambda}\Gamma^{\mu}\left(\unity +
\beta\right)\epsilon\,,  \nonumber\\
\delta_{\epsilon} A_\mu &=& \ft{1}{4}\bar\lambda \Gamma_\mu\big(\unity +
\beta\big)\epsilon\nonumber\\
&&
+\ft{1}{8} \alpha^2 \bar\lambda\Gamma_m (\ft{1}{3} \unity +
\beta)
\epsilon \bar\lambda \Gamma^m \partial_\mu \lambda
+\ft{1}{4}\alpha\bar{\lambda}\Gamma^\rho \left(\unity +
\beta\right)\epsilon F_{\rho\mu}\,,
\label{delepstopdown}
\end{eqnarray}
where $\beta $ is a matrix ($\hat{\Gamma }^\mu =\Pi ^\mu _m\Gamma ^m$)
\begin{eqnarray}
  \beta&=&\left[ \det\left( \delta _\mu {}^\nu +\alpha {\cal F}_{\mu \rho }G^{\rho \nu }\right) \right] ^{-1/2}\sum^{5}_{k=0} \frac{\alpha^k}{2^k k!} \hat \Gamma^{\mu_1
\nu_1 \cdots \mu_k \nu_k }\mathcal{F}_{\mu_1 \nu_1}\cdots
\mathcal{F}_{\mu_k \nu_k}\nonumber\\
&=&1+{\cal O}(\alpha)\,.
 \label{betadefinition}
\end{eqnarray}
Furthermore, there are 16  $\zeta$ transformations:
\begin{eqnarray}
\delta_{\zeta} \lambda &=& \alpha^{-1}\zeta -\ft12\alpha\partial_{\mu}\lambda\bar\lambda \Gamma ^\mu \zeta \,,  \nonumber\\
 \delta_{\zeta} A_\mu&=&-\ft12\bar\lambda \Gamma _\mu \zeta -\ft12\alpha\bar\lambda \Gamma ^\rho  \zeta  F_{\rho\mu}
- \ft{1}{12} \alpha^2 \bar\lambda
\Gamma_m \zeta\bar\lambda \Gamma^m \partial_\mu \lambda\,.
\label{delzetatopdown}
\end{eqnarray}
Note that these transformations do not transform states of a fermion
field to states of a bosonic field, and are thus not regular
supersymmetries. They are transformations of the Volkov--Akulov
(VA)-type. To stress this difference, we say that the theory has $16+16$
supersymmetries.

When we expand the action in orders of $\alpha $, we find that the
action (\ref{Sd10gf}) agrees with (\ref{Sbottomup}) when we choose the
coefficients
\begin{equation}
c_1=2\,,\qquad  c_2=0\,,\qquad  c_3=-1\,,\qquad  c_4=-\ft{1}{2}\,.
\label{preferredci}
\end{equation}
This eliminates in fact all $\partial F$ terms from (\ref{Sbottomup}).

Also the transformation laws (\ref{delepstopdown}) and
(\ref{delzetatopdown}) can be identified with those in the bottom-up
approach, modulo a `zilch symmetry', i.e. a trivial on-shell symmetry.
To complete the identification, $\zeta $ is recognized as a linear
combination of $\epsilon $ and $\eta $.

This proves that our all-order result is indeed the full deformed theory
that we were looking for.

\begin{figure}
\centering
\includegraphics[height=6cm]{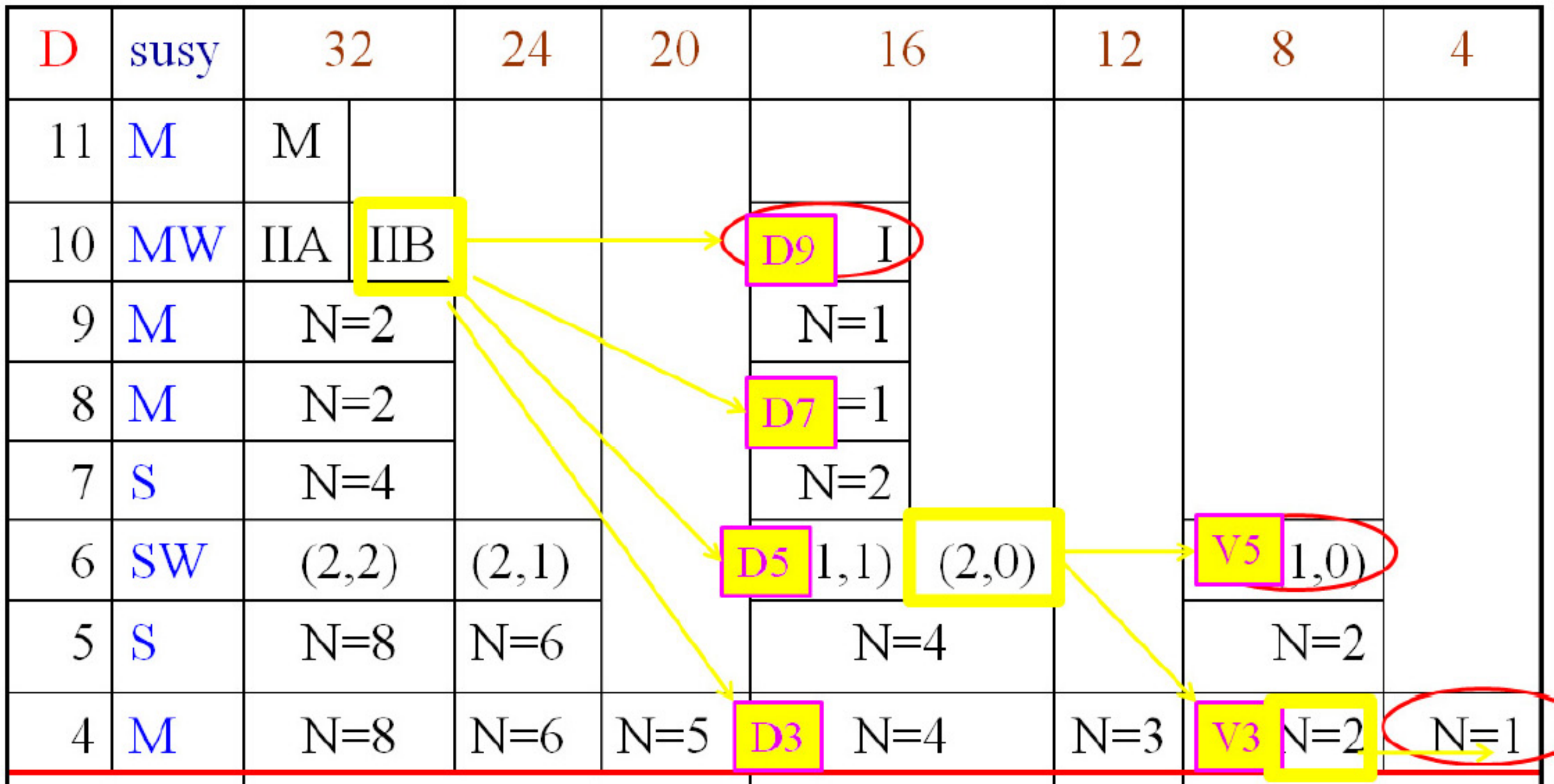}
\caption{The IIB supergravities have solutions denoted as D9,D7,D5,D3. The (2,0) theory has solutions V5,V3  and $N=2$ supergravity has a $N=1$ solution.
The red circles indicate the
basic super-Maxwell theories that we started from in the bottom-up approach and are obtained as maximal $p$ theories. }
\label{fig:allbranes}
\end{figure}

The theories that we can obtain in this way are schematically indicated
in Figure \ref{fig:allbranes}.
The supergravities with each a doublet of local symmetries from which
one starts are indicated as open yellow boxes. The branes type DBI
actions are the D9,D7,D5,D3 when we start from $D=10$, and are indicated
as V5 and V3 when we start from $D=6$. The V stands for vector branes as
explained in the talk of Eric Bergshoeff. This thus shows that we can
construct deformed super-Maxwell theories for various dimensions and
supersymmetry extensions, including $\cn=4$, $\cn=2$ and $\cn=1$ in 4
dimensions.
\subsection{$D=4$, $\cn=4$ gauge multiplet}
Let us in particular consider the D3 case, i.e. the $D=4$, $\cn=4$
theory that we discussed in previous sections. The full action of the
deformed theory is
\begin{equation}
  S =  -\frac{1}{\alpha^2} \int \rmd^{4} x\,\left\{
\sqrt{- \det (G_{\mu\nu} + \alpha {\cal F}_{\mu\nu})} - 1 \right\}\,, \qquad \mu=0,1,2,3\,,
 \label{D3actiongf}
\end{equation}
with
\begin{eqnarray}
G_{\mu\nu} &=& \eta_{m n} \Pi_\mu^{m}
\Pi_\nu^{n}= \eta_{m' n'} \Pi_\mu^{m'}
\Pi_\nu^{n'}+\delta_{IJ} \Pi_\mu^I\Pi_\nu^J \,, \quad  m'=0,1,2,3\,, \nonumber\\
\Pi_\mu^{m'} &=& \delta^{m'}_{\mu}
+\ft12\alpha^2\bar\lambda \Gamma^{m'}
\partial_\mu \lambda \ , \qquad \Pi_\mu^I = \partial_\mu\phi^I
+\ft12\alpha^2\bar\lambda \Gamma^I
\partial_\mu \lambda \,,\quad I=1,...,6 \,,\nonumber\\
 {\cal F}_{\mu\nu}& \equiv& F_{\mu\nu} +\alpha\bar{\lambda}\Gamma_{[\mu }\partial_{\nu]}\lambda
+\alpha\bar{\lambda}\Gamma_{I}\partial_{[\mu}\lambda\partial_{\nu]}
\phi^I\,.
\end{eqnarray}
There are 16 $\epsilon $ and 16 $\zeta $ symmetries, and the remainders
of the rigid Poincar{\'e} transformations in $D=10$ lead to shift symmetries
for the 6 scalars $\phi ^I$. We can compare this with the usual
formulation of the $\cn=4$, $D=4$ super-Maxwell theory:
\begin{equation}
  S_{\rm Maxw}= \int \rmd^4x \Big (-\ft14 F_{\mu\nu} F^{\mu\nu }  -\bar
\psi_i \slashed{\partial} \psi^i -\ft18\partial _\mu\varphi_{ij}
\partial^\mu \varphi^{ij}\Big )\,.
 \label{SMaxwN4D4}
\end{equation}
$F_{\mu \nu }$ is the field strength of the vector field, the $\psi _i$
are 4 Majorana spinors, written as Weyl spinors using the notations
$\psi ^i=\ft12(1+\gamma _*)\psi ^i$ and $\psi _i=\ft12(1-\gamma _*)\psi
_i$. The 6 scalar fields are here represented as antisymmetric tensors
$\varphi_{ij}$, with
\begin{equation}
  \varphi^{ij} \equiv (\varphi_{ij})^*= -\ft12 \varepsilon^{ijk\ell } \varphi_{k\ell }\,.
 \label{varphicc}
\end{equation}
One can find (\ref{SMaxwN4D4}) and the transformation laws as the
$\alpha =0$ part of (\ref{D3actiongf}) and (\ref{delepstopdown}), by
making some identifications. The scalars $\phi ^I$ representing the 6
remaining coordinates in $D=10$ according to (\ref{gaugefixgctkappa})
are divided in two triplets $\phi_a$ and $\phi_{a+3}$ and we identify
\begin{equation}
  \alpha \varphi _{ij}= \phi_a \beta_{ij}^a -{\rm i} \phi_{a+3} \alpha ^a_{ij}\,
,  \qquad a=1,2,3\,,
 \label{identifyvarphi}
\end{equation}
where $\alpha ^a_{ij}$ and $\beta_{ij}^a$ are the Gliozzi--Scherk--Olive
$4\times 4$ matrices \cite{Brink:1976bc,Gliozzi:1976qd}. These are also
used to identify the $D=10$ Majorana-Weyl spinor $\lambda $ introduced
in (\ref{gaugefixgctkappa}), with the 4 Majorana spinors $\psi ^i$. This
is done with the $D=10$ gamma matrix representation
\begin{eqnarray}
 \Gamma ^\mu  & = & \gamma ^\mu \otimes \unity _8 \,,\quad
 \Gamma ^a = \gamma _*\otimes \pmatrix{0&\beta ^a \cr -\beta ^a&0}\,,\quad
 \Gamma ^{a+3}=\gamma _*\otimes \pmatrix{0&\rmi\alpha  ^a \cr \rmi\alpha  ^a&0}\,,\nonumber\\
 C_{10}&=&C_4\otimes\pmatrix{0&\unity _4 \cr \unity _4&0}\,,\qquad
 \Gamma _*=\gamma _*\otimes \pmatrix{\unity _4&0 \cr 0&-\unity _4}\,,
 \label{Gamma10gamma4}
\end{eqnarray}
where $C_{10}$ and $C_4$ are the charge conjugation matrices (for
notation, see \cite{Freedman:2012zz}) in 10 and 4 dimensions, and
$\gamma ^\mu $ are the $D=4$ gamma matrices. In this basis, $\lambda $
is decomposed as
\begin{equation}
  \lambda =\pmatrix{\psi   ^i\cr\psi    _i}\,.
 \label{lambdaaspsi}
\end{equation}
With these identifications, the $\alpha =0$ part of (\ref{D3actiongf})
agrees with (\ref{SMaxwN4D4}). Since the action (\ref{D3actiongf}) is
invariant to all orders in $\alpha $ under the $16+16$ supersymmetries,
it gives the fully consistent deformation of the $\cn=4$, $D=4$ gauge
multiplet. It has both type of supersymmetries: ordinary SUSY and
VA-type supersymmetry. It can be written in the usual 4-dimensional
notations using the translations (\ref{identifyvarphi}) and
(\ref{lambdaaspsi}), but the $D=10$ formulation is much simpler.
\subsection{Worldvolume theory in AdS background}
In order to make progress for $\cn=4$, $D=4$ supergravity, we would need
the deformed gauge multiplet with the superconformal symmetries. The
extra VA symmetries are not of the type of $S$-supersymmetry.
Inspiration may come from old work
\cite{Claus:1997cq,Claus:1998mw,Claus:1998ts} where the worldvolume
theories of branes were considered in an AdS background, leading to a
superconformal theory on the brane. The AdS backgrounds exist only in
particular dimensions and extensions, corresponding to the fact that the
superconformal theories also only exist for particular cases as
explained at the end of Sect. \ref{ss:scmethod}, see Table
\ref{tbl:DNsc}. These actions on the brane are of the form
\begin{eqnarray}
S_{cl}&=&S_{\rm DBI}+S_{\rm WZ}    \,,\nonumber\\
S_{\rm DBI}&=&- \int \rmd^{p+1}  \sigma  \sqrt{- \det
\left( g^{\rm ind}_{\mu\nu}   +F_{\mu\nu}\right) } \,,\nonumber\\
 g^{\rm ind}_{\mu\nu}&=&  \partial_\mu X^{ M} \partial_\nu X^{ N}
G _{ M  N}\,,
\end{eqnarray}
where $G _{ M  N}$ denotes the AdS $\times $ sphere metric that is a
solution of the embedding theory. The theory has then rigid symmetries
inherited from the solution. These are the AdS isometries and the
isometries of the sphere and the corresponding supersymmetries. The
brane theory has as in Sect. \ref{ss:topdown} the worldvolume general
coordinate transformations and kappa symmetries as local symmetries.
After gauge fixing these, the remaining (global) symmetries appear as
conformal symmetries on the brane. The fermionic ones are then $\epsilon
$ ordinary supersymmetry and $\eta $ special supersymmetry. Hence this
is very similar to the appearance of ordinary and VA type
supersymmetries in our new work \cite{Bergshoeff:2013pia}. This gives us
a hope to obtain an all-order deformation of gauge multiplet theories
with superconformal symmetries in the cases where the superalgebras
exist, which includes the D3 brane with $\cn=4$, $D=4$ supersymmetry.

\section{Conclusions}
 \label{ss:concl}

Superconformal symmetry has been used as a tool for constructing
classical actions of supergravity. Also higher-derivative terms can be
constructed with superconformal tensor calculus
\cite{Hanaki:2006pj,deWit:2010za,Bergshoeff:2011xn,Bergshoeff:2012ax,Ozkan:2013uk}.
Quantum calculations show that there are unknown relevant properties of
supergravity theories. We have investigated the possibility that
(broken) superconformal symmetry be such an extra quantum symmetry
\cite{Ferrara:2012ui}. The non-existence of (broken)
superconformal-invariant counterterms and anomalies for $\cn=4$, $D=4$
supergravity could in that case explain 'miraculous' vanishing results.
However, we do not have a systematic knowledge of which
higher-derivative supergravity actions can be invariant under
supersymmetry at all orders in derivatives.

In order to get more insight, we have been looking to gauge multiplets
in global supersymmetry \cite{Bergshoeff:2013pia}. We first considered a
perturbative approach, i.e. constructing actions and transformation laws
order by order in a dimensionful parameter $\alpha $, which can be
related to the string coupling constant. Starting from D$p$ brane
actions in $D=10$ we can construct DBI-type actions that have ordinary
supersymmetry plus VA-type supersymmetry with 16+16 components. They are
related to IIB supergravity, and thus exist for $p=9,7,5,3,...$, leading
to global supersymmetry actions for gauge multiplets in $p+1$
dimensions. For $p=3$ this is the deformation of $\cn=4$, $D=4$ with
higher order derivatives. One can also start from the iib theory in
$D=6$. Also in that case DBI-VA actions (related to objects called
vector branes or `V-branes' \cite{Bergshoeff:2012jb}) with 8+8
supersymmetries. This leads e.g. to the deformation of $D=4$, $\cn=2$
vector multiplets. We hope that insight in these new constructions can
lead also to supergravity actions using the superconformal methods.

\section{Acknowledgements}
Most results in this paper are obtained in collaboration with E.
Bergshoeff, F. Coomans, R. Kallosh, S. Ferrara and  C.~S.~Shahbazi, and
part of the talk has been prepared in collaboration with R. Kallosh. We
thank S. Bellucci and other organizers of this topical meeting, which
was very stimulating.

This work was supported in part by the FWO - Vlaanderen, Project No.
G.0651.11, and in part by the Interuniversity Attraction Poles Programme
initiated by the Belgian Science Policy (P7/37).

\providecommand{\href}[2]{#2}\begingroup\raggedright\endgroup

\end{document}